# Nonmonotonic Temperature-dependent Resistance in Low Density 2D Hole Gases

A. P. Mills, Jr., A. P. Ramirez, L. N. Pfeiffer and K. W. West

*Bell Labs, Lucent Technologies, 600 Mountain Avenue, Murray Hill, NJ 07974*

The low temperature longitudinal resistance-per-square $R_{xx}(T)$ in ungated GaAs/Al$_x$Ga$_{1-x}$As quantum wells of high peak hole mobility $1.7 \times 10^6$ cm$^2$V$^{-1}$s$^{-1}$ is metallic for 2D hole density $p$ as low as $3.8 \times 10^9$ cm$^{-2}$. The electronic contribution to the resistance, $R_{el}(T)$, is a nonmonotonic function of T, exhibiting thermal activation, $R_{el}(T) \propto \exp\{-E_a/kT\}$, for $kT \ll E_F$ and a heretofore unnoted decay $R_{el}(T) \propto 1/T$ for $kT > E_F$. The form of $R_{el}(T)$ is independent of density, indicating a fundamental relationship between the low and high T scattering mechanisms in the metallic state.

The cooperative behavior of fermions confined to two dimensions continues to provide challenges to our understanding of many-body physics. Until recently it had been thought that the T=0 ground state of a 2D fermion gas was an insulator. Kravchenko et al. [1] showed that in high electron mobility Si MOSFET's there is clear evidence for a metal to insulator (MI) transition near $n \approx 10^{11}$ cm$^{-2}$, in contradiction to one parameter scaling theory [2] which predicts that all states in a 2D system should be localized in the limit T→0. The Kravchenko et al. effect was also found using 2D hole systems confined in gated GaAs single interfaces [3,4] driven to sufficiently low hole densities $p \approx 10^{10}$ cm$^{-2}$ where correlation effects are very important. Data on the metallic side of the MI transition has been analyzed in terms of a thermally activated resistance per square [3,5]

$$R_{xx}(T) = R_0 + R_a \exp\{-E_a/kT\}, \qquad (1)$$

with $R_0$ being the residual resistance per square due to impurity scattering.

Despite many suggestions [5-11; see Ref 9 for a complete set of references], there is as yet no consensus on an explanation for the activated resistance. Theories that invoke the effect of remote ionized impurities, for example, through density fluctuations or scattering [9,11] imply a dependence of the activated resistance on the residual resistance. The existence of such a correlation suggests that new insight into the activated resistance mechanism might be gained by studying high mobility samples. In this work we present measurements on GaAs symmetric quantum wells of exceptionally high peak hole-mobility, $\mu_p = 1.7 \times 10^6$ cm$^2$V$^{-1}$s$^{-1}$. We determine $R_{el}(T)$, the electronic contribution to $R_{xx}(T)$, by subtracting the phonon and impurity contributions. In agreement with previous studies, we find low temperature thermally activated behavior. The high temperature behavior is proportional to 1/T and, surprisingly, scales with the low temperature activation energy. This implies a density independent relationship between the low and high temperature scattering mechanisms in the metallic state.

Our samples were made on (311)A GaAs wafers using Al$_x$Ga$_{1-x}$As barriers (typical x=0.10) and symmetrically placed Si delta-doping layers above and below pure GaAs quantum wells of width 30 nm. The samples were prepared with inversion-symmetric potential wells doped from both sides to minimize the effect of spin-orbit coupling which can lift the degeneracy of the two lowest energy hole bands. Pudalov [5] suggested that the asymmetry-induced non-degeneracy might cause a thermally activated resistance effect [12]. For this same reason the hole-density was varied from 0.38 to 8.7 $\times 10^{10}$ cm$^{-2}$ by changing the growth parameters and not by the use of a gate. The samples were prepared in the form of Hallbars, of approximate dimensions (2.5 $\times$ 9) mm$^2$, with diffused In(5%Zn) contacts. The measurement current (~1 nA, 5 Hz) was applied along the [233] direction. Independent measurements of the longitudinal resistance per square, $R_{xx}$, from contacts on both sides of the sample were made simultaneously as the temperature or applied magnetic field was varied. The samples have hole mobilities, extrapolated to T=0, greater than 5 $\times 10^5$ cm$^2$V$^{-1}$s$^{-1}$ and represent to our knowledge the highest found to date in this type of structure. A subset of the Shubnikov-de Haas oscillations of $R_{xx}$ measured at 50 mK to determine p is shown in Fig 1.

Our measurements of $R_{xx}(T)$ are shown in Fig 2. It is evident that the low temperature resistance is nonmonotonic for the lowest density samples and cannot be described by the Arrhenius function of Eq 1. The high temperature data of Fig 2 are dominated by Bloch-Gruneisen phonon scattering, which has an asymptotic T$^3$ dependence at low temperatures and a linear dependence on T at high temperatures. We approximate the phonon contribution by the form [13]

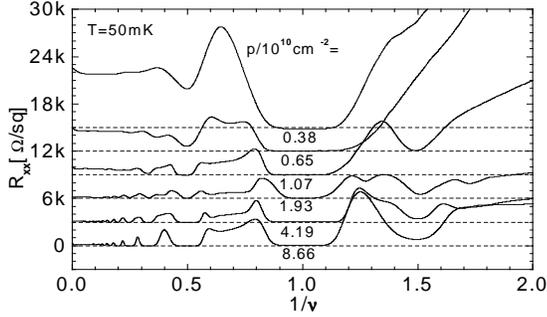

Figure 1. Longitudinal resistance per square vs. magnetic field divided by the field corresponding to the $\nu=1$ Shubnikov-de Haas dip for samples of various hole densities p. The traces are displaced vertically in steps of 3 k$\Omega/\square$.

$R_{ph}t^3/(1+t^2)$, where $t \equiv T/T_{ph}$. The temperature dependence of $R_{xx}$ is thus

$$R_{xx}(T) = R_0 + R_{ph}t^3/(1+t^2) + R_{el}(T) \qquad (2)$$

Having examined a number of simple functions, we found that we could fit our data well using

$$R_{el}(T) = R_a g(x) \exp\{-x^{-1}\},$$

where $g(x) = x^{-1}(\alpha^{-2} + x^{-2})^{-1/2}$, $x \equiv kT/E_a$ and $\alpha$ is a constant. The electronic contribution to the resistance per square is identical to the thermally activated component of Eq(1) at low T, but decreases as $1/T$ at high temperature.

The curves of Eq(2) fitted to the data are shown in Fig 2. In our fit, we find that it is a good approximation to set $\alpha=2.5$, independent of p. The fitted values of $E_a$, $R_a$ and the peak resistance $R_{peak} = 0.477R_a + R_0$ (where the maximum value of $\rho(x)$ is $\rho(2.24) \approx 0.477$) are plotted in Fig 3. The fitted values of the mobility extrapolated to T=0, $\mu = 1/(epR_0)$, and the Bloch-Gruneisen parameters $R_{ph}$ and $T_{ph}$ are plotted in Fig 4. The different, although not orthogonal, shapes of $R_{ph}t^3/(1+t^2)$ and $R_{el}(T)$ combined with the smooth variation of the parameters in Fig 4 suggest that we are able to determine the phonon and impurity scattering contributions to $R_{xx}(T)$ without unduly influencing the fitted shape of $R_{el}(T)$[14].

Having determined the phonon and impurity scattering contributions to the resistance, we subtract them from the data and display in Fig 5 our data in the form

$$[R_{xx}(T) - R_0 + R_{ph}t^3/(1+t^2)]/R_a = \rho(kT/E_a). \qquad (3)$$

Within the limitations imposed by the range of temperatures, the data all fall on a single curve. The form of this dimensionless resistance function is approximately

$$\rho(x) = g(x)\exp\{-x^{-1}\}, \qquad (4)$$

as arrived at by the fitting procedure used to deduce the phonon and impurity contributions to the resistance. It is interesting to note that if the three curves corresponding to the highest density are eliminated from Fig 4, the form of $\rho(x)$ is unchanged. In this case, since $E_a$ is roughly proportional to p for the lower densities (see Fig 3b), the scaling variable is proportional to $E_F$.

The data on our high-mobility samples is consistent with that of Refs. 3 and 15 in the region where we have similar values of $R_0$. We find among our samples a relationship, $R_a \propto R_0^{3/2}$, with the same proportionality constant as for the metallic samples of Ref. 3 [16]. Our activation energy is also consistent with the measurements of Ref. 3 [17]. A significant difference is that the metal-to-insulator transition described in Ref 15 is observed to occur at a critical density $p_c \approx 7.7 \times 10^9 cm^{-2}$, whereas we continue to observe metallic behavior for p as low as $3.8 \times 10^9 cm^{-2}$. The transition in Ref 15 is thus at least partially impurity-driven rather being a disorder-free Wigner-solid transition. In the presence of significant numbers of impurities the density may be less important for driving the metal-insulator transition than is the value of p where the maximum metallic resistance equals $h/e^2$. Extrapolating our fitted peak resistance $R_{peak} \equiv R_0 + 0.477R_a$ vs. p data to low p [see Fig 3a], we find that at $p = (2.3 \pm 0.4) \times 10^9 cm^{-2}$, $R_{peak} \approx$

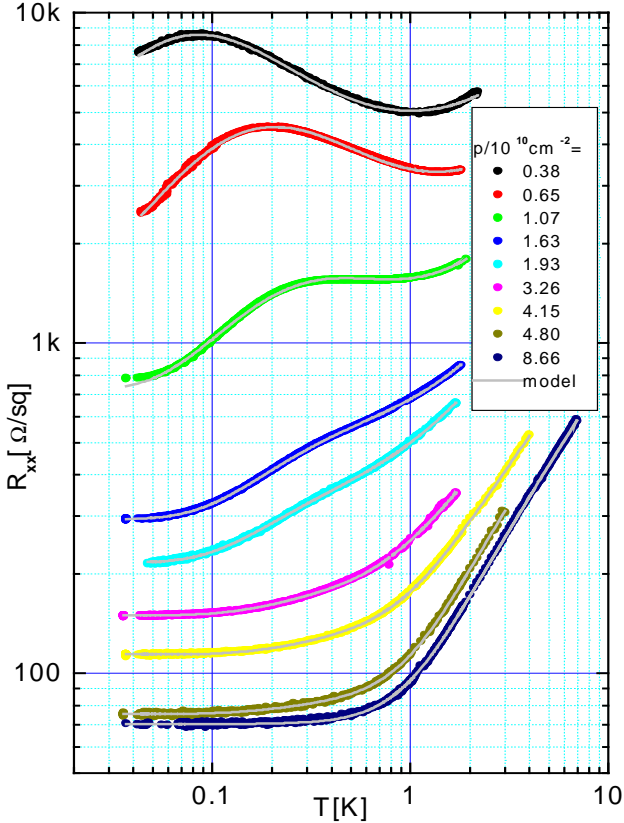

Figure 2. Longitudinal resistance per square vs. temperature for samples of various hole densities p. The model fit of Eq 2 is indicated by the gray lines.

$h/e^2 = 25.8$ k$\Omega$/□. Furthermore, a linear extrapolation [see Fig 3b] suggests that $E_a$ will vanish at $p = (2.0\pm0.4)\times10^9$ cm$^{-2}$. Assuming the vanishing of $E_a$ is a necessary condition at the separatrix between the metallic and insulating phases [18], these two extrapolations are in agreement with a critical density $p_c \approx 2\times10^9$ cm$^{-2}$ for our system. This estimate of $p_c$ is close to the calculated critical density below which the pure Wigner solid forms, $p_c = (1.8+0.6-0.4)\times10^9$ cm$^{-2}$ [19,20].

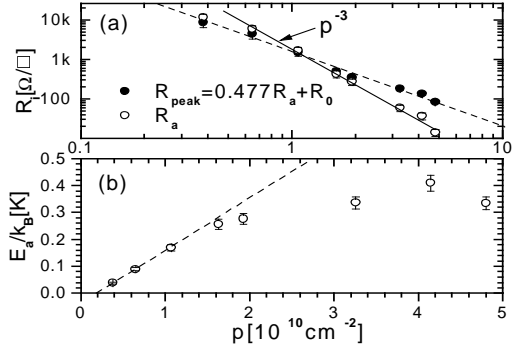

Figure 3. Parameters for the model fit of Eq 2 vs. hole density p. (a) Peak resistance $R_{peak}$ and magnitude of the resistance function $R_a$. (b) Activation energy $E_a$ in units of degrees Kelvin.

The nonmonotonic dimensionless resistance function of Eq. 4 is the main result of this paper. Previous analyses of R(T) near the MI transition in low carrier density samples have focussed on the low-temperature limiting behavior. On the insulating side of $p_c$, R(T) exhibits an activated hopping conductance, whereas on the metallic side, the resistance is activated in the limit T$\rightarrow$0 [18]. Indeed the low-temperature behavior of Eq. 4 is consistent with the scaling behavior found in previous measurements, both on Si-MOSFETS and GaAs heterostructures. The unusual and unexpected result of our measurements is that the scaling behavior persists up to much higher temperatures, including the decrease of R(T) with increasing T. If our resistance function is taken at face value [21], its high temperature behavior provides a natural description of the "tilted separatrix" which has been observed in most of the measurements. Our measurements would then imply that the same scattering process is responsible for the high-T behavior in the metallic state as in the insulating state.

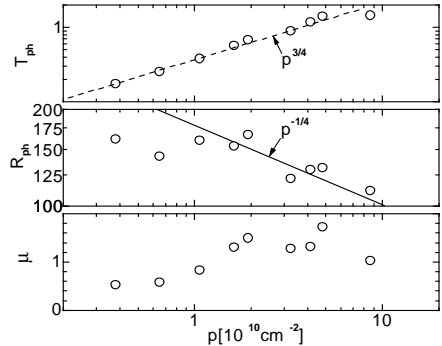

Figure 4. Parameters for the model fit of Eq 2 vs. hole density p. (a) Bloch-Gruneisen temperature $T_{ph}$ [K]. (b) Bloch-Gruneisen resistance $R_{ph}$ [$\Omega$]. (c) Hole mobility [$10^6$ cm$^2$V$^{-1}$s$^{-1}$] deduced from the residual resistance $R_0$ extrapolated to T=0.

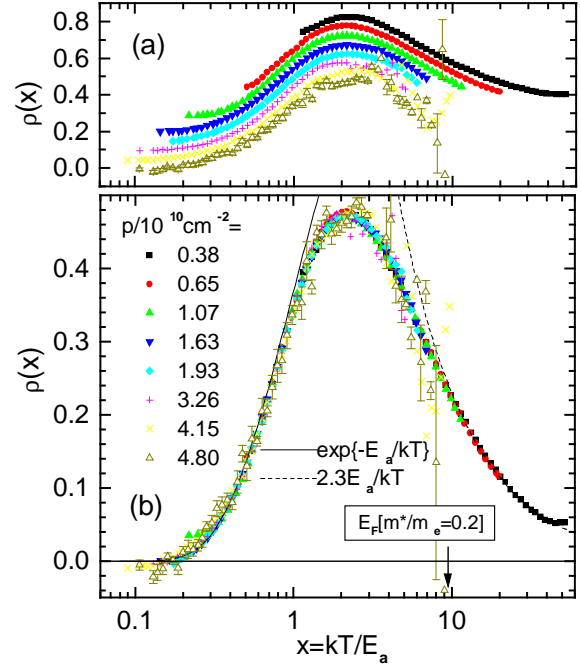

Figure 5. Dimensionless resistance function $\rho(x)$ where $x=kT/E_a$ obtained from Fig 2 according to Eq (2). (a) $\rho(x)$ with data for each density displaced by multiples of 0.05 to exhibit the range of the reduced temperature x. (b) Same as (a) with the data unshifted. The asymptotic behavior of $\rho(x)$ from Eq(4) is shown by the solid and dotted lines. The arrow indicates the approximate Fermi energy for the four lowest densities.

Without some new mechanism, the nonmonotonic R(T) in the metallic state is difficult to explain for a single 2D band. Above $T_F$, classical impurity scattering obtains, yielding R(T) $\propto 1/T$, as is observed and as has already been predicted by Das Sarma and Hwang [11]. However, in the limit of a degenerate Fermi liquid, R(T) should display power-law behavior as T approaches 0 instead of the exponential behavior we and others observe. Other theoretical approaches invoke an additional subsystem in order to explain the MI transition itself and these might be modified to produce a nonmonotonic R(T). Altschuler and Maslov [9] proposed that ionized impurities in the insulating AlGaAs region bind carriers to the side of the 2D region which then cause dipole-scattering of the carriers. The temperature dependence of R(T) in this model arises from the activation of the impurity charge-state. A deficiency of the model relative to our experiment is the constancy of R(T) at high temperature. The theory by

Finkelshtein [22] which combines weak localization ideas with strong Coulomb interactions predicts a nomonotonic R(T), but without quantitative agreement with our observations. Another possibility is that the nonmonotonicity of R(T) arises from a collective effect related to transient or localized regions of partially-formed Wigner solid [7].

In conclusion, we have measured the low temperature longitudinal resistance in ungated hole-doped GaAs/Al$_x$Ga$_{1-x}$As quantum wells of exceptionally high peak mobility $1.7 \times 10^6$ cm$^2$V$^{-1}$s$^{-1}$. The resistance is metallic for 2D hole density p as low as $3.8 \times 10^9$ cm$^{-2}$ and is in excess of the contributions attributable to impurity and phonon scattering, as previously observed. However, the functional form of the excess resistance is not simply activated, but is nonmonotonic and independent of density, exhibiting thermal activation at low temperature and 1/T decay at high temperature. This is evidence for a nontrivial scattering mechanism in the metallic state.

The authors are pleased to thank E. Abrahams, B. L. Altshuler, P. M. Platzman, S. H. Simon, D. C. Tsui, A. Turberfield and C. M. Varma for helpful discussions.

———

[1] S. V. Kravchenko, D. Simonian, M. P. Sarachik, W. Mason and J. E. Furneaux, *Phys. Rev. Lett.* **77**, 4938 (1996).
[2] E. Abrahams, P. W. Anderson, D. C. Licciardello and T. V. Ramakrishnan, *Phys. Rev. Lett.* **42**, 673 (1979).
[3] Y. Hanein, U. Meirav, D. Shahar, C. C. Li, D. C. Tsui and H. Shtrikman, *Phys. Rev. Lett.* **80**, 1288 (1998).
[4] M. Y. Simmons, A. R. Hamilton, M. Pepper, E. H. Linfield, P. D. Rose and D. A. Ritchie, *Phys. Rev. Lett.* **80**, 1292 (1998).
[5] V. M. Pudalov, *JEPT Lett.* **66**, 175 (1997).
[6] V. Dobrosavljevic, E. Abrahams, E. Miranda and S. Chakravarty, *Phys. Rev. Lett.* **79**, 455 (1997).
[7] S. He and X. C. Xie, *Phys. Rev. Lett.* **80**, 3324 (1998).
[8] C. Castellani, C. Di Castro and P. A. Lee, *Phys. Rev. B* **57**, R9381 (1998).
[9] B. L. Altshuler and D. L. Maslov, *Phys. Rev. Lett.* **82**, 145 (1999).
[10] S. V. Kravchenko, M. P. Sarachik and D. Simonian, Comment on "Theory of metal-insulator transitions in gated semiconductors", cond-mat/9901005.
[11] S. Das Sarma and E. H. Hwang, "Charged impurity scattering limited low temperature resistivity of low density silicon inversion layers", cond-mat/98122116.
[12] S. J. Papadakis, E. P. De Poortere, H. C. Manoharan, M. Shayegan and R. Winkler, *Science* **283**, 2056 (1999), have observed an activated resistance associated with an asymmetry applied to an otherwise symmetric quantum well via a transverse electric field. This could be a confirmation of the effect envisioned by Pudalov [5]. If so, the activation energy associated with the residual asymmetry in our samples would be one to two orders of magnitude smaller than our measured effect and have a p$^2$ dependence not in agreement with our measurements [see our Fig 3b].
[13] V. Karpus, *Semi. Sci. Techn.* **5**, 691 (1990). According to this author, the crossover temperature T$_{ph}$ should be proportional to p$^{3/4}$ and R$_{ph}$ should be proportional to p$^{-1/4}$, independent of the residual resistance R$_0 \equiv R_{xx}(0)$.
[14] At the lowest three densities where the phonon contribution to R$_{xx}$ is becoming negligible over much of the temperature range, T$_{ph}$ could not be determined independently of R$_{ph}$ and was scaled as p$^{0.75}$ from the fitted values at higher densities [13]. Also at the lower three densities, R$_{ph}$ exhibits a slower variation than the p$^{-0.25}$ suggested by Ref 13.
[15] J. Yoon, C. C. Li, D. Shahar, D. C. Tsui, and M. Shayegan, *Phys. Rev. Lett.*, **82**, 1744 (1999).
[16] To make this comparison, we have digitized the data of Ref 3 Fig 1 and fitted it using our Eq 2.
[17] Hanein et al. [3] find that E$_a$ is proportional to p over the range of their measurements. Our well width w=30 nm is such that at high densities, the Coulomb repulsion across the well causes the hole liquid to have some of the features of a double well. In this case the dielectric function will differ substantially from that of the single 2D carrier sheet that describes carriers in a single interface. This effect could explain the nonlinearity of E$_a$(p) exhibited in Fig 3(b).
[18] V. M. Pudalov, G. Brunthaler, A. Prinz and G. Bauer, *JETP Lett* **68**, 442 (1998).
[19] B. Tanatar and D. M. Ceperley, *Phys. Rev. B* **39**, 5005 (1989), calculate that the Wigner crystallization threshold density is r$_s$=37±5.
[20] From B. E. Cole, et al., *Phys. Rev. B* **55**, 2503 (1997), the zero field limit of the cyclotron resonance mass for p=5×10$^{10}$ cm$^{-2}$ in a 15 nm quantum well doped from one side is m*=(0.18±0.01)m$_e$. We thank Prof. Tsui for bringing this reference to our attention. Taking this as the appropriate bare hole band mass for comparison to the calculation of Tanatar and Ceperley [19], the dimensionless Wigner-Seitz radius for our system is r$_s$=(15.5±0.9)×(10$^{10}$cm$^{-2}$/p)$^{1/2}$.
[21] It may be necessary to modify your resistance function by allowing the parameter α to be density dependent as the density is lowered below the critical density.
[22] A. M. Finkelshtein, *Sov. Phys. JETP* **57**, 97 (1983).